\documentclass[twocolumn,english,aps,prb,showpacs]{revtex4}
\usepackage[T1]{fontenc}
\usepackage[latin9]{inputenc}
\usepackage{amsmath}
\usepackage{amssymb}
\usepackage{graphicx}

\makeatletter
\@ifundefined{textcolor}{}
{%
 \definecolor{BLACK}{gray}{0}
 \definecolor{WHITE}{gray}{1}
 \definecolor{RED}{rgb}{1,0,0}
 \definecolor{GREEN}{rgb}{0,1,0}
 \definecolor{BLUE}{rgb}{0,0,1}
 \definecolor{CYAN}{cmyk}{1,0,0,0}
 \definecolor{MAGENTA}{cmyk}{0,1,0,0}
 \definecolor{YELLOW}{cmyk}{0,0,1,0}
 }
\usepackage{babel}

\usepackage{epsfig}\usepackage{subfigure}\usepackage{amsfonts}\usepackage{CJK}\@ifundefined{definecolor}
 {\usepackage{color}}{}
\usepackage{graphics}
\makeatother

\begin{document}

\title{Analog Superconducting Quantum Simulator for Holstein Polarons}

\author{Feng Mei$^{1}$, Vladimir M. Stojanovi\'{c}$^{2,3}$, Irfan Siddiqi$^{4}$, and Lin Tian$^{1}$}

\affiliation{$^{1}$School of Natural Sciences, University of California, Merced, CA 95343, USA\\
$^{2}$Department of Physics, University of Basel, Klingelbergstrasse 82, CH-4056 Basel, Switzerland\\
$^{3}$Department of Physics, Harvard University, Cambridge, MA 02138, USA\\
$^{4}$Quantum Nanoelectronics Laboratory, Department of Physics, University of California, Berkeley, CA 94720, USA}

\begin{abstract}
We propose an analog quantum simulator for the Holstein molecular-crystal model based on a superconducting circuit QED system in the dispersive regime. By varying the driving field on the superconducting resonators, one can readily access both the adiabatic and anti-adiabatic regimes of this model. Strong e-ph coupling required for small-polaron formation can also be reached. We show that small-polaron state of arbitrary quasimomentum can be generated by applying a microwave pulse to the resonators. We also show that significant squeezing in the resonator modes can be achieved in the polaron-crossover regime through a measurement-based scheme. 
\end{abstract}
\maketitle

\section{introduction\label{sec:intro}}
Quantum simulation of many-body systems opens up an exciting perspective for studying condensed matter and high energy effects that cannot be studied by traditional theoretical or experimental techniques.~\cite{Feynman, Lloyd} Owing to recent progress in quantum devices, the realization of quantum simulators for a broad spectrum of problems, such as quantum magnetism and  quantum Hall effects, has been intensively studied.~\cite{Cirac+Zoller:12} At the same time, the questions of how to exploit the unique features of each specific physical system to probe and manipulate the many-body state and the dynamics of the simulator remain to be answered.

The Holstein molecular-crystal model is commonly used to study the short-range coupling between fermionic excitation (electron, hole) and optical phonons (e-ph coupling).~\cite{Holstein} The coupling in this model has the form of a local interaction between the fermion density and the lattice displacement, and has important consequences on the optical and transport properties of the solids.~\cite{PolaronMaterial} One of the most fundamental many-body effects due to this coupling is the formation of a small polaron where an extra charge carrier becomes heavily dressed in a cloud of virtual phonons of the host crystal.~\cite{PolaronPhysics} The Holstein model does not admit analytical solution and can only be solved approximately by numerical methods. A quantum simulator for this model can advance our understanding of the behavior of polaronic systems. This simple many-body system can also give us hands-on experience in effectively manipulating quantum simulators built from a specific architecture. In previous works, simulators for the Holstein- and related models were proposed with cold polar molecules~\cite{HerreraEtAl} and trapped ions.~\cite{Stojanovic+:12,Mezzacapo+:12} However, the accessible parameter regimes and the effectiveness of these simulators are limited by intrinsic physical and technical constrains in these systems. 

The flexibility and control of superconducting (SC) quantum circuits provide us with an excellent platform for quantum simulation.~\cite{squbitReview, squbitExp} It was shown that quantum spin systems can be simulated with SC qubits that have demonstrated ever increasing coherence times.~\cite{squbitSpinSimu} SC resonators are ideal for simulating bosonic degrees of freedoms such as phonons. The strong qubit-resonator coupling  demonstrated in circuit quantum electrodynamics (circuit QED) experiments~\cite{CQEDtheory, CQEDexp} adds a Hubbard-like interaction for the microwave photons in the resonators and can be used to study quantum phase transitions in such systems.~\cite{CQEDSimuReview, CQEDSimuRecent} The diversity of the SC devices also enables the simulation of complex quantum processes such as universal quantum computation and exciton transport.~\cite{squbitSimuOther} Here, we propose an analog SC quantum simulator for the one-dimensional Holstein model. The central building block of our simulator is a circuit QED system composed of a transmon qubit and a SC resonator and operated in the dispersive regime.~\cite{KochSchoelkopfPRA2007TransmonQubit} The role of the qubits is to simulate fermionic excitations and the resonator modes almost perfectly mimic Einstein phonons. The only tunable parameters required for accessing both the adiabatic and anti-adiabatic regimes and for preparing a small-polaron state of arbitrary quasimomentum are the amplitude and frequency of the microwave drive on the resonators. The coupling strength required for small-polaron formation can be readily reached. A striking feature of this simulator is that measurement-based squeezing up to $1.25\,\textrm{dB}$ in the resonator modes can be achieved in the polaron state in the crossover regime. Meanwhile, detection of the polaron states can be achieved through an ancilla qubit (probe qubit) that couples with one of the resonator modes. Compared with previous proposals for the Holstein model,~\cite{HerreraEtAl, Stojanovic+:12,Mezzacapo+:12} our proposal effectively simulates this model with essentially dispersionless phonons and hopping processes via Josephson couplings which naturally have nearest-neighbor character. 

This paper is organized as the following. In Sec.~\ref{sec:simulator}, we present a circuit-QED-based superconducting quantum simulator for the Holstein model and derive the many-body Hamiltonian for this system. By applying the Jordan-Wigner transformation, this Hamiltonian can be exactly mapped to the Holstein model. The accessible parameter regimes of this simulator are studied in Sec.~\ref{sec:crossover}. Using a variational method, we show that small-polaron formation under strong e-ph coupling can be achieved with practical circuit parameters. In Sec.~\ref{sec:preparation}, we present a scheme that can prepare the simulator state into a polaron state with arbitrary quasimomentum. The anomalous amplitude fluctuation and momentum squeezing in the polaron ground state are studied in Sec.~\ref{sec:squeezing}. In Sec.~\ref{sec:discussions}, we discuss the detection of the polaron state. We also study the effects of decoherence and quantum leakage on the quantum simulator. Conclusions are given in Sec.~\ref{sec:conclusions}. 

\section{The simulator\label{sec:simulator}} 
\begin{figure}
\includegraphics[width=8.5cm,clip]{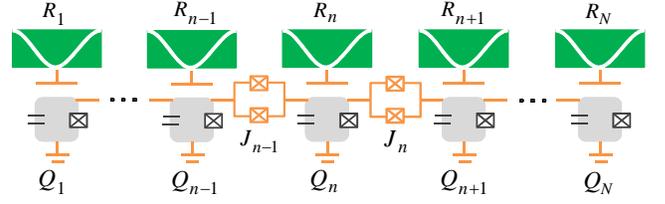} 
\caption{(Color online) Schematic setup of the SC simulator for the Holstein model with the transmon qubits denoted by $Q_{n}$, SC resonators denoted by $R_{n}$, and SQUID loops denoted by $J_{n}$.}
\label{fig:circuit}
\end{figure}

The repeating unit of this simulator is made of a transmon qubit denoted by $Q_{n}$ capacitively coupled with a SC resonator denoted by $R_{n}$, as is shown in Fig.~\ref{fig:circuit}. The resonators can be in various forms such as coplanar waveguide or lumped element resonators. The Hamiltonian of the repeating unit is described by the Jaynes-Cummings model
\begin{equation}
H_{0}^{n}=\hbar\omega_{c}a_{n}^{\dagger}a_{n}+\frac{\hbar\omega_{z}}{2}\sigma_{n}^{z}+\hbar g\left(a_{n}^{\dagger}\sigma_{n}^{-}+\sigma_{n}^{+}a_{n}\right),\label{eq:Hn}
\end{equation}
where $\omega_{c}$ and $\omega_{z}$ are the frequencies of the resonator and qubit respectively, $g$ is the magnitude of the qubit-resonator coupling, $a_{n}$ is the annihilation operator of the resonator mode, and $\sigma_{n}^{z,\pm}$ are the Pauli operators of the qubit. Adjacent qubits couple via a SQUID loop denoted by $J_{n}$ with effective Josephson energy $E_{J}$. The coupling Hamiltonian is $H_{J}^{n}=-E_{J} \cos(\varphi_{n}-\varphi_{n+1})$ in terms of the gauge-invariant phases.~\cite{OrlandoBook1991} For transmon qubits, we can write
\begin{equation}
H_{J}^{n}\approx-t_{0}(\sigma_{n}^{+}\sigma_{n+1}^{-}+\sigma_{n+1}^{+}\sigma_{n}^{-})\label{eq:HJn}
\end{equation}
with hopping matrix element $t_{0}=E_{J}\delta\phi_{0}^{2}$ and quantum displacement $\delta\phi_{0}$ of the phase variables (see Appendix A for details). In addition, the resonators are driven by a microwave source which is described by the Hamiltonian 
\begin{equation}
H_{d}^{n}=2\varepsilon_{0} \cos(\omega_{d}t) (a_{n}+a_{n}^{\dagger})\label{eq:Hdn}
\end{equation}
with driving amplitude $\varepsilon_{0}$ and driving frequency $\omega_{d}$. The total Hamiltonian of this simulator is hence $H_{t} =\sum_{n} (H_{0}^{n}+H_{J}^{n}+H_{d}^{n})$.

In the dispersive regime of $|\Delta|\gg g$ with $\Delta\equiv\omega_{c}-\omega_{z}$ being the qubit-resonator detuning, we apply the unitary  transformation 
\begin{equation}
U=\prod_{n}e^{-\frac{g}{\Delta}(\sigma_{n}^{+}a_{n}-a_{n}^{\dagger}\sigma_{n}^{-})}\label{eq:U}
\end{equation}
to the simulator Hamiltonian.~\cite{CQEDtheory} The term $H_{0}^{n}$ is transformed into
\begin{equation}
\bar{H}_{0}^{n}=\hbar\omega_{c}a_{n}^{\dagger}a_{n}+\frac{\hbar}{2}(\omega_{z}-\chi)\sigma_{n}^{z}-\hbar\chi\sigma_{n}^{z}a_{n}^{\dagger}a_{n},\label{eq:Hnbar}
\end{equation}
with the Stark shift $\chi\equiv g^{2}/\Delta$. The terms $H_{J}^{n}$ and $H_{d}^{n}$ are also transformed accordingly. In the interaction picture and after a displacement of the resonator modes $(a_{n}\rightarrow a_{n}-\varepsilon_{0}/\hbar \delta\omega)$, the total Hamiltonian becomes
\begin{equation}
\bar{H}_{r}=\sum_{n} \hbar\delta\omega \left[ a_{n}^{\dagger}a_{n}+g_{H}\frac{\sigma_{n}^{z}+1}{2}\left(a_{n}+a_{n}^{\dagger}\right)\right] +H_{J}^{n}\label{eq:Hrbar}
\end{equation}
with $\delta\omega\equiv\omega_{c}+\chi-\omega_{d}$ and $g_{H}\delta\omega=2\varepsilon_{0}\chi/\hbar \delta\omega$. Note that we assume $\varepsilon_{0}\gg\hbar \delta\omega$ in deriving this Hamiltonian. Details of the derivation of the above Hamiltonian can be found in Appendix A. 

By applying the Jordan-Wigner transformation ($\sigma_{n}^{+}=c_{n}^{\dag}\prod_{m=1}^{n-1}e^{i\pi c_{m}^{\dag}c_{m}}$ and $\sigma_{n}^{z}=2 c_{n}^{\dag}c_{n}-1$), we derive
\begin{equation}
\bar{H}_{r}=\sum_{n}  \hbar \delta\omega \left[ a_{n}^{\dagger}a_{n}+g_{H} c_{n}^{\dagger}c_{n}\left(a_{n}+a_{n}^{\dagger}\right)\right] +\bar{H}_{J}^{n}\label{HolsteinHam}
\end{equation}
with $\bar{H}_{J}^{n}=-t_{0}(c_{n}^{\dagger}c_{n+1}+c_{n+1}^{\dag}c_{n})$ and $c_{n}$ being the annihilation operator of the fermionic excitations at site $n$. This Hamiltonian has the standard form of the Holstein model with $\delta\omega$, $t_{0}$ and $g_{H}$ playing the roles of phonon frequency, nearest-neighbor hopping matrix element, and dimensionless e-ph coupling, respectively. Note that given the diversity of SC circuits, other types of SC qubits such as the flux qubit can also be used to construct a quantum simulator for the Holstein model. In Appendix B, we present a flux-qubit-based quantum simulator for this model. 

\section{Polaron crossover\label{sec:crossover}} 
\begin{figure}
\includegraphics[width=8.5cm,clip]{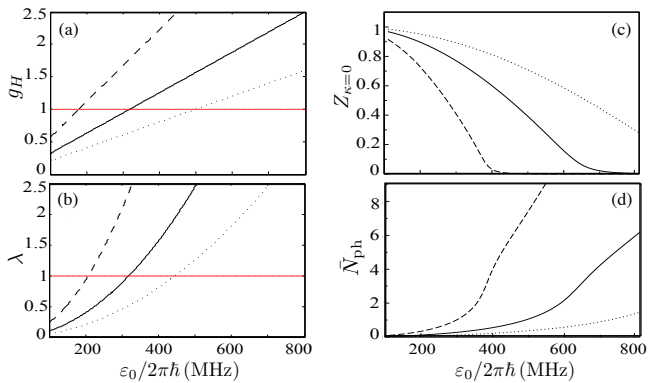}
\caption{Dimensionless coupling constants (a) $g_{H}$ and (b) $\lambda$, (c) Quasiparticle residue $Z_{\kappa=0}$, and (d) mean phonon number $\bar{N}_{ph}$ versus driving amplitude $\varepsilon_{0}$. The dashed, solid, and dotted curves are for $\hbar\delta\omega/t_{0}=0.75,\,1,\,1.25$, respectively.}
\label{fig:polaron}
\end{figure}

By varying the driving parameters ($\varepsilon_{0}$, $\omega_{d}$), all interesting regimes of the Holstein model can be accessed where a fermonic excitation displays qualitatively different behavior. The adiabatic (anti-adiabatic) regime can be accessed by choosing $\hbar \delta\omega/t_{0}$ to be smaller (larger) than one. With $\lambda= g_{H}^{2}\hbar \delta\omega/t_{0}$, the conditions for small-polaron formation are $g_{H},\,\lambda >1$. In Fig.\ref{fig:polaron} (a) and (b), we plot $g_{H}$ and $\lambda$ at selected $\delta\omega$ values for a practical set of parameters: $g/2\pi=200\,\textrm{MHz}$, $\Delta/2\pi=4\,\text{GHz}$, and $t_{0}/2\pi\hbar=80\,\textrm{MHz}$. It can be seen that the crossover from quasi-free excitation to strongly-dressed small-polaron state can be realized in both the adiabatic and anti-adiabatic regimes. For example, at $\varepsilon_{0}/2\pi\hbar=400\,\textrm{MHz}$ and $\delta\omega/2\pi=80\,\textrm{MHz}$, we obtain $g_{H}=1.25$ and $\lambda=1.56$. Note that an optional control on the simulator is to tune the hopping matrix element $t_{0}$ by applying a global magnetic flux to the SQUID loops $J_{n}$, which can adjust the adiabaticity of the system. 

To demonstrate the polaron crossover in the simulator, we apply a variational method to find the small-polaron ground state using the Toyozawa Ansatz which provides a rather accurate estimate of the ground-state energy of the Holstein model in all relevant physical regimes.~\cite{Toyozawa:61} The ground state can be obtained by minimizing the energy expectation value with respect to the variational parameters in the Ansatz (see Appendix C for details). One important quantity for characterizing polaron excitation is the quasiparticle residue $Z_{\kappa}\equiv|\langle\Psi_{k=\kappa}|\tilde{\psi}_{\kappa}\rangle|^{2}$ which is defined as the overlap between the dressed polaron state $\vert\tilde{\psi}_{\kappa}\rangle$ at quasimomentum $\kappa$ and the bare-excitation Bloch state $|\Psi_{k}\rangle\equiv c_{k}^{\dagger}|0\rangle$ at momentum $k=\kappa$ with $c_{k}=\sum c_{n}e^{ikn}/\sqrt{N}$. This quantity can characterize the crossover from the bare-excitation regime into the small-polaron regime. Another quantity characterizing the polaron crossover is the mean phonon number $\bar{N}_{\text{ph}}\equiv{\langle\tilde{\psi}_{\kappa=0}|\:\sum_{i}a_{i}^{\dagger}a_{i}\:|\tilde{\psi}_{\kappa=0}\rangle}$ in the polaron ground state. For the Toyozawa Ansatz, both $Z_{\kappa}$ and $\bar{N}_{\textrm{ph}}$ can be evaluated in terms of the optimal values of the variational parameters. In Fig.\ref{fig:polaron} (c) and (d), these two quantities are shown using the parameter values given above for a system size of $N=32$. The change of $Z_{\kappa=0}$ from unity to values very close to zero as $\varepsilon_{0}$ increases is a clear manifestation of the smooth crossover from a quasi-free excitation to a small polaron state. The same crossover is also illustrated by the mean phonon number which varies from nearly zero to $\bar{N}_{\text{ph}}\gtrsim3$ as $\varepsilon_{0}$ increases. Note that we also calculate the above quantities for small systems of, e.g., $N=4$ (which is easier to realize in experiments), and find nearly identical results due to the local nature of the lattice distortion in the small-polaron regime.

\section{Polaron-state preparation\label{sec:preparation}} 
To study polaron crossover, extra fermionic excitation needs to be prepared in the simulator. Without this extra excitation, the many-body state can be written as $\left\vert G_{0} \right\rangle =\left\vert 0\right\rangle _{\textrm{e}}\otimes\left\vert 0\right\rangle _{\textrm{ph}}$ in the excitation-phonon basis. In the physical basis of the qubit-resonator system, this state has all the qubits in the spin down state and all the resonators in the vacuum state. In the SC circuit, this state can be prepared via thermalization in a low temperature environment. Here we show that given the initial state $\left\vert G_{0}\right\rangle$, a small-polaron state with arbitrary quasimomentum can be generated through a qubit-flip scheme.

Consider applying a pumping pulse on the resonators in the form of 
\begin{equation}
H_{p}=\varepsilon_{p}(t)\sum(a_{n}^{\dag}e^{-iqn}+a_{n}e^{iqn})/\sqrt{N}\label{eq:Hp}
\end{equation}
with time-dependent driving amplitude $\varepsilon_{p}(t)$ and wave vector $q$. After applying the transformation $U$ given in Eq.(\ref{eq:U}) (with $Ua_{n}U^{\dag}\approx a_{n}-(g/\Delta)\sigma_{n}^{-}$),~\cite{CQEDtheory} we obtain an effective pumping Hamiltonian on the qubits
\begin{equation}
\Omega(q, t)=\beta(t)\sum_{n}\left(\sigma_{n}^{+}e^{-iqn}+\sigma_{n}^{-}e^{iqn}\right)/\sqrt{N},\label{eq:Drive}
\end{equation}
with $\beta(t)=-(g/\Delta)\varepsilon_{p}(t)$, which describes a spin-flip operation on the qubits with site-dependent factor $e^{-iqn}$.  After the Jordan-Wigner transformation, we apply this operator to the state $\vert G_{0}\rangle$. With $c_{n}\left\vert G_{0}\right\rangle =0$, we find that $\Omega(q, t)\vert G_{0}\rangle=\hbar \beta(t)c_{q}^{\dag}\vert G_{0}\rangle$, generating a bare excitation of momentum $q$. Hence the transition matrix element can be written as $\langle\tilde{\psi}_{\kappa}\vert\Omega(q, t)\vert G_{0}\rangle=\hbar \beta(t)\Omega_{q\kappa}$ with $\Omega_{q\kappa}= \langle \tilde{\psi}_{\kappa}\vert c_{q}^{\dag} \vert G_{0}\rangle$. Using the lattice translational symmetry of the system, we derive that
\begin{equation}
\left|\Omega_{q\kappa}\right|=\sqrt{Z_{\kappa}}\delta_{q,\kappa},\label{eq:Omegaqk}
\end{equation}
yielding nonzero matrix element only for $q=\kappa$. 

Let $\hbar\omega_{p}$ be equal to the energy difference between the state $\vert G_{0}\rangle$ and the polaron state $\vert\psi_{\kappa}\rangle$. By choosing $q=\kappa$ and $\beta(t)=2\beta_{p}\cos(\omega_{p}t)$, and under the rotating wave approximation, the pumping in Eq.(\ref{eq:Drive}) generates a Rabi oscillation between the initial state $\vert G_{0}\rangle$ and the target state $\vert\tilde{\psi}_{\kappa}\rangle$. This oscillation is governed by the effective Hamiltonian 
\begin{equation}
\bar{H}_{p}=\beta_{p}\left(\Omega_{\kappa\kappa}\left\vert \tilde{\psi}_{\kappa}\right\rangle \left\langle G_{0}\right\vert +\Omega_{\kappa\kappa}^{*}\left\vert G_{0}\right\rangle \left\langle \tilde{\psi}_{\kappa}\right\vert \right)\label{eq:Hp}
\end{equation}
with a Rabi frequency $\beta_{p}|\Omega_{\kappa\kappa}|$. Starting from the state $\vert G_{0}\rangle$, the system evolves to the target state $\vert\tilde{\psi}_{\kappa}\rangle$ in a duration $\tau=\pi\hbar/(2\beta_{p}\sqrt{Z_{\kappa}})$. For the polaron state $\vert\tilde{\psi}_{\kappa=0}\rangle$, $Z_{\kappa=0}$ decreases with the increase of $\varepsilon_{0}$ as is shown in Fig.\ref{fig:polaron} (c) and it takes a longer time to generate a strongly-dressed polaron state. For $\beta_{p}/2\pi\hbar=20\,\textrm{MHz}$ and $Z_{\kappa}=0.7$, the state preparation time is $\tau=18\,\textrm{\text{ns}}$. This corresponds to a practical value of $\varepsilon_{p}/2\pi\hbar=400\,\textrm{MHz}$ with the parameters given previously. 

In this process, the condition $q=\kappa$ ensures momentum conservation and the choice of the pumping frequency ensures energy conservation. This scheme can be generalized and applied to quantum simulators for other many-body systems to generate elementary excitations by exploiting the symmetry in these systems. 

\section{Anomalous fluctuation and squeezing\label{sec:squeezing}} 
\begin{figure}
\includegraphics[width=8.5cm,clip]{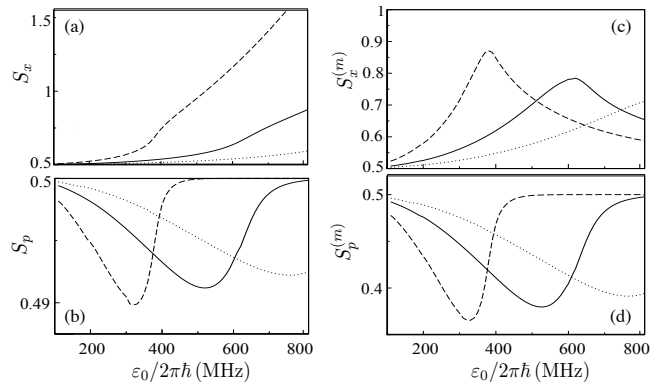}
\caption{The variances of the resonator modes (a) $S_{x}$, (b) $S_{p}$, (c) $S_{x}^{(m)}$, and (d) $S_{p}^{(m)}$ versus $\varepsilon_{0}$. The dashed, solid, and dotted curves are for $\hbar\delta\omega/t_{0}=0.75,\,1,\,1.25\,\textrm{MHz}$, respectively.}
\label{fig:variance}
\end{figure}

In the Holstein model, the interplay between the strong e-ph coupling and the hopping of the fermionic excitation can induce anomalous fluctuation in the phonon modes. In our simulator, this fluctuation occurs when the driving amplitude $\varepsilon_{0}$ increases to reach the crossover regime. We denote the variance of an operator $A$ by $S_{A}=\langle(A-\langle A\rangle)^{2}\rangle$. This quantity characterizes the fluctuation of the operator around its average value. For the position quadrature $x_{n}\equiv(a_{n}+a_{n}^{\dagger})/\sqrt{2}$ and the momentum quadrature $p_{n}\equiv-i(a_{n}-a_{n}^{\dagger})/\sqrt{2}$ of mode $a_{n}$, their variances $S_{x}$ and $S_{p}$ are shown in Fig.\ref{fig:variance} (a) and (b). The variance $S_{x}$ is always larger than the quantum limit of $1/2$ and increases monotonically as $\varepsilon_{0}$ increases, which clearly demonstrates the crossover to the small-polaron regime. The variance $S_{p}$ varies in a very narrow region below $1/2$ with the product $S_{x}S_{p}>1/4$, reflecting the non-Gaussian nature of the fluctuation in the small-polaron state. Due to the lattice translational symmetry, these variances do not depend on the site index $n$. 
 
The above variances can be viewed as the averaged fluctuation of the resonator modes by tracing out the fermionic excitation. Below we study the variances of a single resonator mode when the fermionic excitation is pinned at this site. Consider the measurement-based position and momentum quadratures 
\begin{align}
x^{(m)}\equiv&\sum c_{n}^{\dagger}c_{n}(a_{n}+a_{n}^{\dagger})/\sqrt{2}; \label{eq:xm}\\ 
p^{(m)}\equiv&-i\sum c_{n}^{\dagger}c_{n}(a_{n}-a_{n}^{\dagger})/\sqrt{2}\label{eq:pm}.
\end{align}
The variances of these quadratures $S_{x}^{(m)}$ and $S_{p}^{(m)}$ describe the fluctuation of the resonator mode $a_{n}$ when the excitation (qubit-flip) is detected at this site. In Fig.\ref{fig:variance} (c) and (d), it can be seen that $S_{x}^{(m)}>1/2$ and $S_{p}^{(m)}<1/2$, a property they share with $S_{x}$ and $S_{p}$. However, as $\varepsilon_{0}$ increases, $S_{x}^{(m)}$ behaves very differently from $S_{x}$ and shows an optimal value in the crossover regime. More interestingly, the momentum quadrature $S_{p}^{(m)}$ can reach a low value of $0.35$, i.e., the post-selected momentum quadrature of $a_{n}$ can be squeezed by up to $1.25\,\textrm{dB}$ when the polaron is detected at this site. For a finite array of $N$ sites, the probability of measuring the polaron excitation at a single site is $1/N$. Our numerical results show that this behavior can be observed in a small array of only, e.g., $N=4$ sites, with a probability of $1/4$, which can be readily realized with current technology. Hence, accessing the crossover regime in the simulator is not only a crucial requirement to study the polaron formation, but also presents us with a novel approach to generate squeezing in the microwave photon modes of the resonators.~\cite{resonatorsqueezing}

\section{Detection and decoherence\label{sec:discussions}} 
A crucial step in the quantum simulation of a many-body system is the detection of the many-body state. In a quantum simulator for the Holstein model, we can characterize the polaron crossover by measuring the mean phonon number $\bar{N}_{ph}$ in the polaron ground state. Because of the lattice translational symmetry, this can be further simplified to the measurement of the mean phonon number of one of the resonators, e.g., $a_{1}$. For this purpose, we add an ancilla qubit $\sigma_{d}$ which couples to the resonator $a_{1}$ only during the measurement. The coupling is in the form of Eq.(\ref{eq:Hn}). When the qubit is far detuned from the resonator mode, the mean phonon number of $a_{1}$ can be obtained by measuring the Stark shift of the qubit. Note that in a different regime when the qubit is in resonance with the resonator mode, the mean phonon number can also be obtained by measuring the qubit. Besides the measurement of the ancilla qubit, in order to achieve the measurement-based squeezing in one of the resonator modes, measurement of the qubit at the same site is required. 

The extra excitation in the simulator corresponds to a flipping of the qubit states and a displacement of the resonator modes which are subject to the decoherence of the qubits or the resonators. The coherence properties of SC qubits and resonators have improved significantly over the past few years. Decoherence time of transmon qubits coupling to on-chip resonators can now reach $10 - 40\,\mu$s.~\cite{squbitdecoherence} For coplanar waveguide resonators, the damping time of the microwave photons can reach the same order of magnitude with a quality factor of $Q=10^{6}$.~\cite{resonatorQfactor} In our simulator, the effective phonon frequency, the e-ph coupling, and the hopping element are all of hundreds of megahertz, far exceeding these decoherence rates. The duration of the state-preparation pulse is several orders of magnitude shorter than the decoherence times. In addition, thermal excitations can be neglected in the low temperature environment as the energy of the excitation is of a few gigahertz. The pump pulses may induce leakage (unwanted transitions) to higher energy levels in the transmon qubit.~\cite{KochSchoelkopfPRA2007TransmonQubit} However, typical anharmonicity of a transmon qubit gives us an off resonance of around $500$ MHz for the unwanted transitions. For a driving amplitude $\beta_{p}/2\pi=25\,\text{MHz}$, the probability of leakage is well below one percent, which is a tolerable error rate for the simulator. 

\section{Conclusions\label{sec:conclusions}} 
To conclude, we propose a circuit QED-based quantum simulator for the Holstein-polaron model. By varying the driving on the resonators, all relevant physical regimes of the Holstein model can be accessed, and in particular, we can reach the strong coupling regime for small-polaron formation. We also show that polaron state of arbitrary quasimomentum can be prepared by pumping the resonators. The polaron state in the crossover regime shows the striking feature of measurement-based squeezing in the resonator modes. Our work not only opens a promising route to study the electron-phonon physics with SC quantum simulators, but can also advance the  control and detection methods for the many-body states in SC simulators by exploiting the unique controllability of such devices.

\section*{ACKNOWLEDGEMENTS\label{sec:acknowledgements}} We thank Christoph Bruder for very helpful discussions. F. M. and L. T. were supported by NSF-DMR-0956064 and NSF-CCF-0916303. V. M. S. was supported by the SNSF and the NCCR QSIT. I. S. acknowledges partial support from the NSF Award 0939514. This research was supported in part by NSF PHY11-25915 through a KITP program.

\section*{APPENDIX A: DERIVATION OF THE EFFECTIVE HAMILTONIAN\label{sec:appendixa}}
The total Hamiltonian of the simulator is given by $H_{t} =\sum_{n} (H_{0}^{n}+H_{J}^{n}+H_{d}^{n})$, where the terms $H_{0}^{n}$, $H_{J}^{n}$, and $H_{d}^{n}$ are given in Eqs.(\ref{eq:Hn}-\ref{eq:Hdn}). In the Josephson coupling $H_{J}^{n}=-E_{J} \cos(\varphi_{n}-\varphi_{n+1})$ between adjacent qubits, the effective Josephson energy of the SQUID loop can be written as $E_{J} = 2E_{J0} \cos(\pi\Phi_{x} / \Phi_{0})$, where $E_{J0}$ is the Josephson energy of the single junctions in the SQUID loop, $\Phi_{x}$ is the static magnetic flux in the loop, and $\Phi_{0}$ is the flux quantum.~\cite{OrlandoBook1991} By adjusting the magnetic flux in the SQUID loop, the effective Josephson energy can be manipulated. For transmon qubits, $H_{J}^{n}$ can be approximated as
\begin{equation}
H_{J}^{n}\approx E_{J}\delta\phi_{0}^{2}\left[ \frac{\sigma_{n}^{z}+\sigma_{n+1}^{z}}{2}- \left(\sigma_{n}^{+}\sigma_{n+1}^{-}+\sigma_{n+1}^{+}\sigma_{n}^{-}\right)\right],\label{eq:HJdetail}
\end{equation}
where we have neglected the constant terms. Here, $\delta\phi_{0}$ is the quantum displacement of the phase variable $\varphi_{n}$ with $\delta\phi_{0}^{2}\sim\sqrt{2E_{C1}/E_{J1}}$ written in terms of the charging energy $E_{C1}$ and the Josephson energy $E_{J1}$ of the transmon qubit. With typical parameters, $\delta\phi_{0}^{2}\sim0.15$~\cite{KochSchoelkopfPRA2007TransmonQubit}. The first term in Eq.\eqref{eq:HJdetail} can be absorbed into the qubit energy $\omega_{z}$. Hence, the Josephson coupling can be simplified as $H_{J}^{n}=-t_{0}\left(\sigma_{n}^{+}\sigma_{n+1}^{-}+\sigma_{n+1}^{+}\sigma_{n}^{-}\right)$ with hopping matrix element $t_{0}=E_{J}\delta\phi_{0}^{2}$. 

Our simulator is operated in the dispersive regime where the magnitude of the detuning far exceeds the magnitude of the qubit-resonator coupling, i.e., $|\Delta|\gg g$. Here the detuning is defined as $\Delta\equiv\omega_{c}-\omega_{z}$. We start by applying the unitary transformation $U$ given in Eq.(\ref{eq:U}) to the total Hamiltonian $H_{t}$. The term $H_{0}^{n}$ is transformed into $\bar{H}_{0}^{n}=UH_{0}^{n}U^{\dagger}$ given in Eq.(\ref{eq:Hnbar}) to the lowest order of the small ratio $g/\Delta$ with $\chi\equiv g^{2}/\Delta$ being the Stark shift. In a similar manner, we can derive the expressions for $UH_{J}^{n}U^{\dagger}$ and $UH_{d}^{n}U^{\dagger}$ and derive the transformed total Hamiltonian $\bar{H}_{t}=UH_{t}U^{\dagger}$. 

For the transformed Hamiltonian, we consider the interaction picture defined by the non-interacting Hamiltonian
\begin{equation}
H_{0}=\sum_{n}\left[ \hbar\omega_{d}a_{n}^{\dagger}a_{n}+(\hbar\bar{\omega}_{z}/2)\sigma_{n}^{z}\right],\label{eq:H0}
\end{equation}
where $\bar{\omega}_{z}=\omega_{z}-\chi-2\chi(\varepsilon_{0}/\hbar \delta\omega)^{2}$ is the modified qubit frequency and $\delta\omega=\omega_{c}+\chi-\omega_{d}$ is the modified resonator detuning. The modified qubit frequency $\bar{\omega}_{z}$ includes the Stark shift and another term that is used to balance the effect of the microwave driving. By applying the rotating wave approximation (RWA) and omitting the fast-rotating terms, the transformed Hamiltonian in the interaction picture can be written as
\begin{equation}
\bar{H}_{r1}=\sum_{n}\left[ H_{1}^{n}+H_{J}^{n}+\varepsilon_{0}\left(a_{n}+a_{n}^{\dagger}\right)\right].\label{eq:Hr0}
\end{equation}
with the term
\begin{equation}
H_{1}^{n}=\hbar \delta\omega a_{n}^{\dagger}a_{n}+\hbar\chi\left(\frac{\varepsilon_{0}}{\hbar \delta\omega}\right)^{2}\sigma_{n}^{z}-\hbar\chi\left(\sigma_{n}^{z}+1\right)a_{n}^{\dagger}a_{n}.\label{eq:H1n}
\end{equation}
Fast-rotating terms such as $a_{n}^{\dagger}\sigma_{n}^{-}$ generated by the transformation $U$ have been omitted under the RWA. Next, we apply a displacement operator~\cite{WallsMilburnBook2006} to shift the resonator modes with $a_{n}\rightarrow a_{n}-\varepsilon_{0}/\hbar \delta\omega$. After this shift, the Hamiltonian $\bar{H}_{r1}$ becomes $\bar{H}_{r}$ in Eq.(\ref{eq:Hrbar}) with the coupling constant $g_{H}\delta\omega=2\varepsilon_{0}\chi/\hbar \delta\omega$. With $\varepsilon_{0}\gg\hbar \delta\omega$, the term $-\hbar\chi \left(\sigma_{n}^{z}+1\right)a_{n}^{\dagger}a_{n}$ has been neglected from the above Hamiltonian. To convert the qubit modes to fermionic excitations, we apply the Jordan-Wigner transformation to the spin operators. The Hamiltonian $\bar{H}_{r}$ then recovers the form in Eq.(\ref{HolsteinHam}).

\section*{APPENDIX B: REALIZATION WITH FLUX QUBIT\label{sec:appendixb}}
Given the diversity of the SC circuits, quantum simulator for the Holstein-like model can also be realized with other SC qubits such as the flux qubit and the phase qubit. Here we present a realization of the Holstein model with the flux qubit.~\cite{fluxqubit} We will show that a total Hamiltonian of the form $H_{t} =\sum_{n} (H_{0}^{n}+H_{J}^{n}+H_{d}^{n})$ can be constructed with the flux qubit.

Consider a flux qubit biased at the degeneracy point (with a bias magnetic flux of $\Phi_{ex}=0.5\Phi_{0}$). The qubit Hamiltonian can be written as  $\hbar\omega_{z}\sigma_{n}^{z}/2$ in terms of the eigenstates, where $\hbar\omega_{z}$ is equal to the quantum tunneling between the two persistent-current states of the flux qubit and the eigenstates are $90$ degrees rotated from the persistent-current states.~\cite{fluxqubit} It was shown in recent experiments that the quantum tunneling can exceed a few gigahertz.~\cite{MITexp2011} The microwave mode of the SC resonator couples to the flux qubit through its magnetic field which inductively couples to the current loop of the qubit and generates a coupling $g(a_{n}+a_{n}^{\dag})\sigma_{n}^{x}$. The magnitude of the coupling can be engineered in a very wide range and can readily reach sub-gigahertz.~\cite{strongcoupling2010} The single-site Hamiltonian can hence be written as Eq.(\ref{eq:Hn}) under the RWA. 

The neighboring qubits naturally couple via their mutual inductance. The coupling Hamiltonian can be written as $H_{J}^{n}=-t_{0}\sigma_{n}^{x}\sigma_{n+1}^{x}$, where $t_{0}=MI_{cir}^{2}$ with $M$ being the mutual inductance between the qubits and $I_{cir}$ being the magnitude of the circulating current of the qubit states. Under the RWA, the coupling can be written as $H_{J}^{n}=-t_{0}\left(\sigma_{n}^{+}\sigma_{n+1}^{-}+\sigma_{n+1}^{+}\sigma_{n}^{-}\right)$ after neglecting the fast-rotating terms. One drawback of this coupling is its long-range nature, which induces coupling between qubits that are not immediately adjacent to each other. However, as the mutual inductance decreases as $1/r^{3}$ with $r$ being an effective distance between two qubits, the coupling between non-neighboring qubits also decreases as $1/r^{3}$. An alternative coupling scheme is to design a tunable coupling between neighboring qubits, where the coupling can be controlled by external sources.

The resonators can be driven by a microwave source in the form of $H_{d}^{n}$. Hence, combining all three terms: $H_{0}^{n}$, $H_{J}^{n}$, and $H_{d}^{n}$, we obtain a total Hamiltonian $H_{t}$ with the flux qubit. Following the procedure presented in Appendix A, we can construct the Holstein model from this Hamiltonian. It can be shown that all relevant physical regimes can also be accessed in this realization.

\section*{APPENDIX C: TOYOZAWA ANSATZ\label{sec:appendixc}}
To determine the polaron ground state of our system, we make use of a variational method which yields results that agree well with quantum Monte Carlo and exact-diagonalization results. As the eigenstates of the Holstein Hamiltonian are good quasimomentum states, the variational states are Bloch-type states $|\psi_{\kappa}\rangle=N^{-1/2}\sum_{n}e^{-i\kappa n}|\psi_{\kappa}(n)\rangle$, where $|\psi_{\kappa}(n)\rangle$ denotes a Wannier-like function of the coupled e-ph system and $\kappa$ is an eigenstate of the total quasimomentum operator $K=\sum_{k}k\:c^{\dagger}_{k}c_{k}+\sum_{q}q\:a^{\dagger}_{q}a_{q}$. Here we use the Toyozawa Ansatz state as our variational state.~\cite{Toyozawa:61} This Ansatz state is given by
\begin{equation}
|\psi_{\kappa}(n)\rangle=\sum_{m=-N/2}^{N/2-1}\Phi_{\kappa}(m)e^{-i\kappa m}c_{n+m}^{\dag}|0\rangle_{\textrm{e}}|\xi_{\kappa}(n) \rangle_{\textrm{ph}}, \label{ToyozawaAnsatz}
\end{equation}
where $|\xi_{\kappa}(n)\rangle_{\mathrm{ph}}\equiv\prod_{l}\exp\big(v_{l}^{\kappa}a_{n+l}^{\dagger}-v_{l}^{\kappa\ast}a_{n+l}\big)|0 \rangle_ {\mathrm{ph}}$ is a direct product of phonon coherent states at sites $n+l$ ($l=-N/2,\ldots,N/2-1$) and the $2N$ variational parameters $\{\Phi_{\kappa}(m),v_{l}^{\kappa}\}$ are complex valued. Here $c_{n}$'s ($a_{m}$'s) are the real space operators of the fermionic excitations (phonons) at site $n$ ($m$). This Ansatz provides a rather accurate estimate of the polaron ground-state energy of the Holstein model in all relevant physical regimes. The ground state can be obtained by minimizing the expectation value $\langle{\psi}_{\kappa=0}|H|{\psi}_{\kappa=0}\rangle/\langle{\psi}_{\kappa=0}|{\psi}_{\kappa=0} \rangle$ with respect to the variational parameters. We also introduce the normalized dressed excitation state at quasimomentum $\kappa$ as $\vert\tilde{\psi}_{\kappa}\rangle=\vert\psi_{\kappa}\rangle/\sqrt{\left\langle \psi_{\kappa}\vert\psi_{\kappa}\right\rangle }$. In the main paper, we use the wave function $\vert\tilde{\psi}_{\kappa}\rangle$ in all our discussions.

\end{document}